\documentclass[final, 5p, times,authoryear, twocolumn]{elsarticle}

\usepackage{amsmath}
\usepackage{siunitx}
\usepackage{graphicx}
\usepackage[dvipsnames]{xcolor}
\usepackage{url}
\usepackage{booktabs}
\usepackage{multirow}
\usepackage[referable]{threeparttablex}
\usepackage{enumitem}

\newcommand{\tabincell}[2]{\begin{tabular}{@{}#1@{}}#2\end{tabular}}

\def\opartial#1#2{{\frac{\partial #1}{\partial #2}}}

\def\diff1#1#2{{\frac{{\rm d} #1}{{\rm d} #2}}}
\def\diffi#1#2#3{{\frac{{\rm d}^#3 #1}{{\rm d} #2^#3}}}

\setcitestyle{square}
\makeatletter
\def\ps@pprintTitle{%
 \let\@oddhead\@empty
 \let\@evenhead\@empty
 \def\@oddfoot{}%
 \let\@evenfoot\@oddfoot}
\makeatother
\begin{document}

\bibliographystyle{model1-num-names}

\begin{frontmatter}

\title{\huge The {GPU}-based Phase Tracking method\\ for Planetary Radio Science applications\tnoteref{t1}}
\tnotetext[t1]{This work has been supported by the National Natural Science Foundation of China, the Astronomical Joint Program (U1531104).}

\author[1]{Nianchuan Jian}
\author[2]{Dmitry Mikushin}
\author[3]{Jianguo Yan\corref{cor}} \cortext[cor]{Corresponding author} \ead{jgyan@whu.edu.cn}
\author[3,4]{Jean-Pierre Barriot}
\author[1]{Yajun Wu}

\address[1]{Shanghai Astronomical Observatory, Chinese Academy of Science, Shanghai 200030, China}
\address[2]{Applied Parallel Computing LLC, Switzerland}
\address[3]{State Key Laboratory of Information Engineering in Surveying, Mapping and Remote Sensing,Wuhan University, Wuhan, 430070, China}
\address[4]{University of French Polynesia, Geodesy Observatory of Tahiti, Laboratoire GEPASUD, BP 6570, 98702 Faaa airport, Tahiti, French Polynesia}

\begin{abstract}
This paper introduces a phase tracking method for planetary radio science research with computational algorithm implemented fo r NVIDIA GPUs. In contrast to the phase-locked loop (PPL) phase counting method used in traditional Doppler data processing, this method fits the tracking data signal into the shape expressed by the Taylor polynomial with optimal phase and amplitude coefficients. The Differential Evolution (DE) algorithm is employed for polynomial fitting. In order to cope with high computational intensity of the proposed phase tracking method, the graphics processing units (GPUs) are employed. As a result, the method estimates the instantaneous phase, frequency, derivative of frequency (line-of-sight acceleration) and the total count phase of different integration scales. This data can be further used in planetary radio science research to analyze the planetary occultation and gravitational fields. The method has been tested on MEX (Mars Express, ESA) and Chang'E 4 relay satellite (China) tracking data. In a real experiment with 400K data block size and $\sim$80,000 DE solver objective function evaluations we were able to acheive the target convergence threshold in 6.5 seconds and do real-time processing on NVIDIA GTX580 and 2$\times$ NVIDIA K80 GPUs, respectively. \textcolor{gray}{The precision of integral Doppler (60s) is 2 mrad/s and 4 mrad/s for MEX(3-way) and Chang'E 4 relay satellite(3-way) respectively.}

\end{abstract}


\end{frontmatter}

\section{Introduction}
    In deep space exploration missions, radio measurements are often used to study natural phenomena of solar system such as mass and gravity of planets, media around planets and internal structure of planets, etc. This type of research is generally called radio science~\citep{asmar1993deep}. Physical parameters of electromagnetic signal, which links station and spacecraft will be changing during its propagation due to the relative motion of spacecraft and station, influence of medias and the different gravitational potential at the position of spacecraft and station. Physical parameters include phase, frequency, amplitude and polarization. The variation of these parameters allows to estimate the atmospheric and ionospheric structure of planets, solar corona, planetary gravitational fields, shapes, mass, planetary rings and ephemerides of planets. Furthermore, parameters variation allows to verify the theory of general relativity and gravitational waves detection and gravitational red shift~\citep{asmar1993deep,asmar2009planetary}. In planetary atmospheric and ionospheric structure studies, the parameters of interest are instantaneous amplitude, phase and frequency. For planetary gravity, ephemerides and the theory of general relativity, the parameters of interest are the integral Doppler or the total count phase. All these parameters can be retrieved through high-precision Doppler measurements. 

Since the late 1960s, the Jet Propulsion Laboratory (JPL) has been conducting radio science research in early deep space explorations. Remarkable results have been achieved in the course of 50 years: the high-precision gravity field of Moon and Mars have been determined by Doppler tracking~\citep{lorell1968lunar,konopliv1998improved,konopliv2006global,lemoine2001improved}; the atmospheric component and ionospheric models of Mars and Venus have been studied from the radio science tracking data. These achievements form the basis of further deep space explorations of human being.

    \subsection{Planetary radio science hardware}
        The radio science research is supported by the NASA Deep Space Network (DSN) located at Goldstone (USA), Madrid (Spain) and Canberra (Australia). The DSN is equipped with large antennas (26, 34, and 70 meters)~\citep{asmar1993deep}, radio receiver and Doppler processor. The fifth generation of JPL Doppler processors was designed in the form of the the so-called ``block series''. The block series are mainly used in precise orbit determination, planetary gravity recovery and general relativity research with close-loop tracking model. The open-loop tracking model receivers have been developed and applied in 1990s' space missions, such as RSR (radio science receiver) by JPL ~\citep{Sniffin2005dsms} and IFMS (Intermediate Frequency Modulation System) by ESA~\citep{patzold2004mars}. The open tracking model is mainly used in atmospheric and ionospheric studies. Recently, some attempts have been made to apply the open tracking model to the planetary gravity research~\citep{highsmith2002effect}.  

In Doppler data processing, the amount of computations is large and requires high-performance analysis. The Application Specific Integrated Circuit (ASIC) technology is employed in design of traditional Doppler extractors to compute the phase accumulation (also called phase counting)~\citep{aung1994block}. Nowadays, the general-purpose computing devices also offer enough performance to be used in Doppler data processing software, e.g. the Universal Software Radio Peripheral (USRP) based on GNU Radio\footnote{\url{https://www.gnuradio.org/}}. The USRP can process Doppler data using the FFT-based methods~\citep{lofaldli2016implementation}.

This paper introduces a new phase tracking method, which employs graphic processing units (GPUs) to compute phase, frequency and frequency derivative of tracking data. The new algorithm fits Taylor polynomial series expanded at data block center to signal existing in the baseband data block\footnote{The data block is usually 2 seconds and sampling is 200KHz.}. The method calculates the analytical form of phase expression in the neighborhood of the data block center and the amplitude with slope of signal. The amount of baseband fitting computations is large and is offloaded onto GPUs. We show that real-time data processing can be acheived by using two NVIDIA Tesla K80 GPUs. The fitting is performed using Differential Evolution (DE) algorithm~\citep{storn1995differrential}~–~a well-known robust method for global optimization problems. From the analytical form of phase we can further deduce the frequency, derivative of frequency (line-of-sight acceleration) and the integral phase (total count phase) of different time scales~\citep{moyer2005formulation}. This data can be used in planetary radio science research to analyze planetary occultation and gravitational fields.

The proposed method has three main advantages over existing hardware and software Doppler processing solutions. Firstly, through the adjustment of the polynomial order and data block length, the analytical form of phase (frequency) will have the phase noise level precision of $\sigma_{p_T}$\footnote{The truncated error is equal to $\sigma_{p_T}$ by adjusting. For details please refer to the error analysis below.} ($T$ is the data length) along the whole block length. In particular, the method will be more accurate in Doppler processing with high sampling rates ($\sim$100Hz), required in planetary occultation research. Secondly, the method can resolve the frequency derivative with precision of $\sigma_{\dot{f}_T}$, directly reflecting the line-of-sight acceleration of the spacecraft relatively to the station (see ~\ref{sec:app3}). Through the frequency derivative we can solve some further dynamic parameters by means of least square fitting. Thirdly, GPUs are less expensive and easier for software development, compared to traditional ASIC-based Doppler processing equipment.

    \subsection{Differential evolution algorithm}
        Doppler data block fitting is a computationally expensive problem, as each objective function evaluation is an integration of time series. Fortunately, the integration can be easily parallelized. Moreover, additional degree of parallelism may be offered by certain kinds of optimization algorithms, such as Differential Evolution (DE). Our numerical results showed that a single objective function evaluation already saturates the compute power of a modern GPU. Therefore, a parallel DE could be performed on multiple GPUs, but we leave this topic out for a separate project. 

The DE is a robust global optimization method widely used in various fields to solve multidimensional problems~\citep{price2006differential}. The DE algorithm originates from genetic annealing algorithm first described by Price et al.~\citep{price1994stimulation}. The DE algorithm proved itself by winning the International Contest on Evolutionary Optimization (ICEO) twice in 1996 and 1997~\citep{storn1996minimizing, price1997differential}.

\begin{figure}[!h]
\centering
\includegraphics[width=8cm,clip=true,height=8cm]{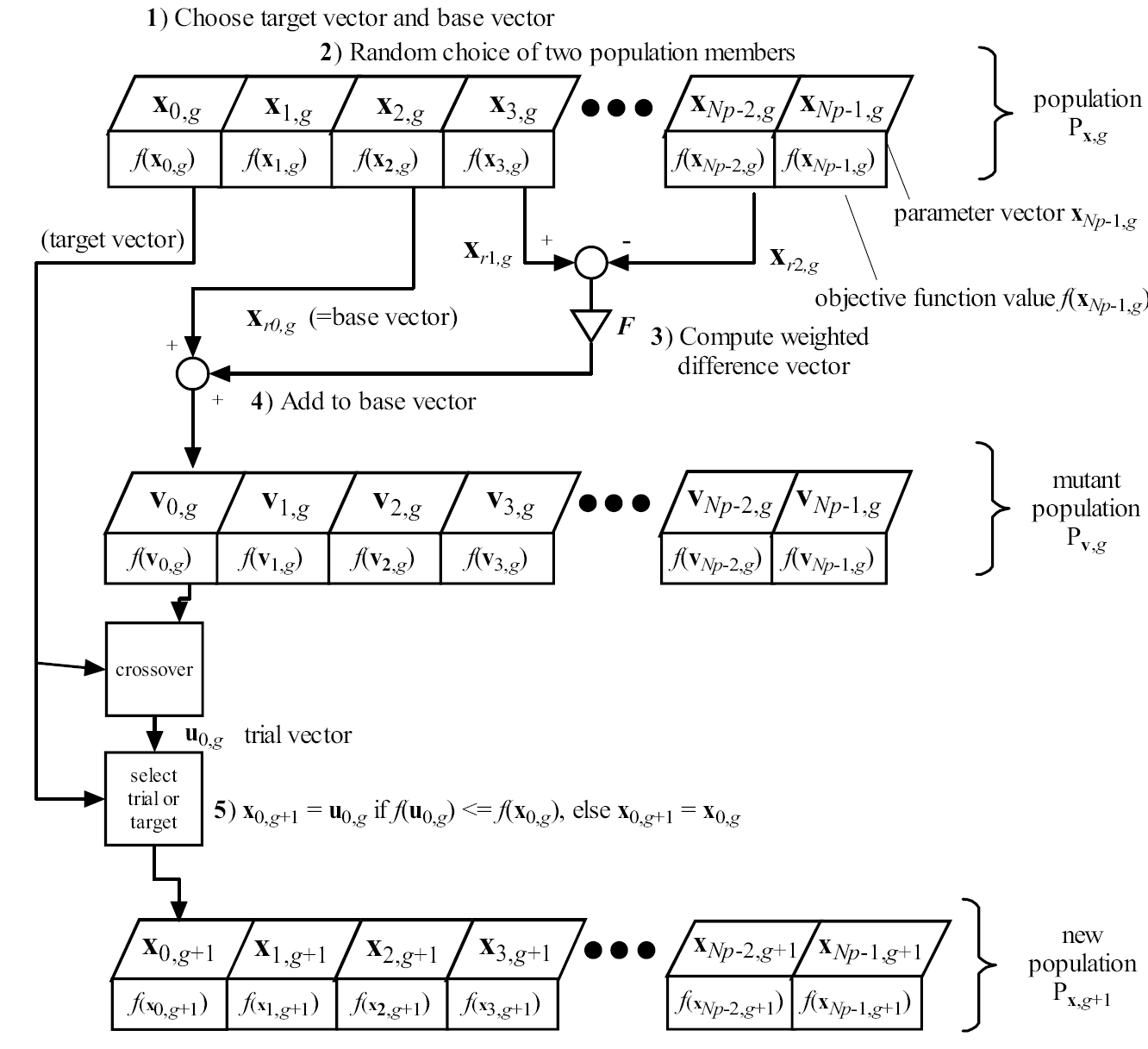}
\caption{DE algorithm iteration flowchart}
\vspace*{-3mm}
\label{fig:de}
\end{figure}

In DE, the initial population is randomly chosen within the parameter space, and the child parameters vectors are iteratively derived from their parents through the ``mutation'' procedure, depending on the objective function response (Fig.~\ref{fig:de}). The mutation is followed by the ``crossover`` path selection. In order to escape from local minima, DE introduces a weighting factor. The iterative process continues until the fitfulment of the convergence criteria.

    \subsection{GPGPU computing}
\begin{figure}[!h]
\centering
\includegraphics[width=9cm]{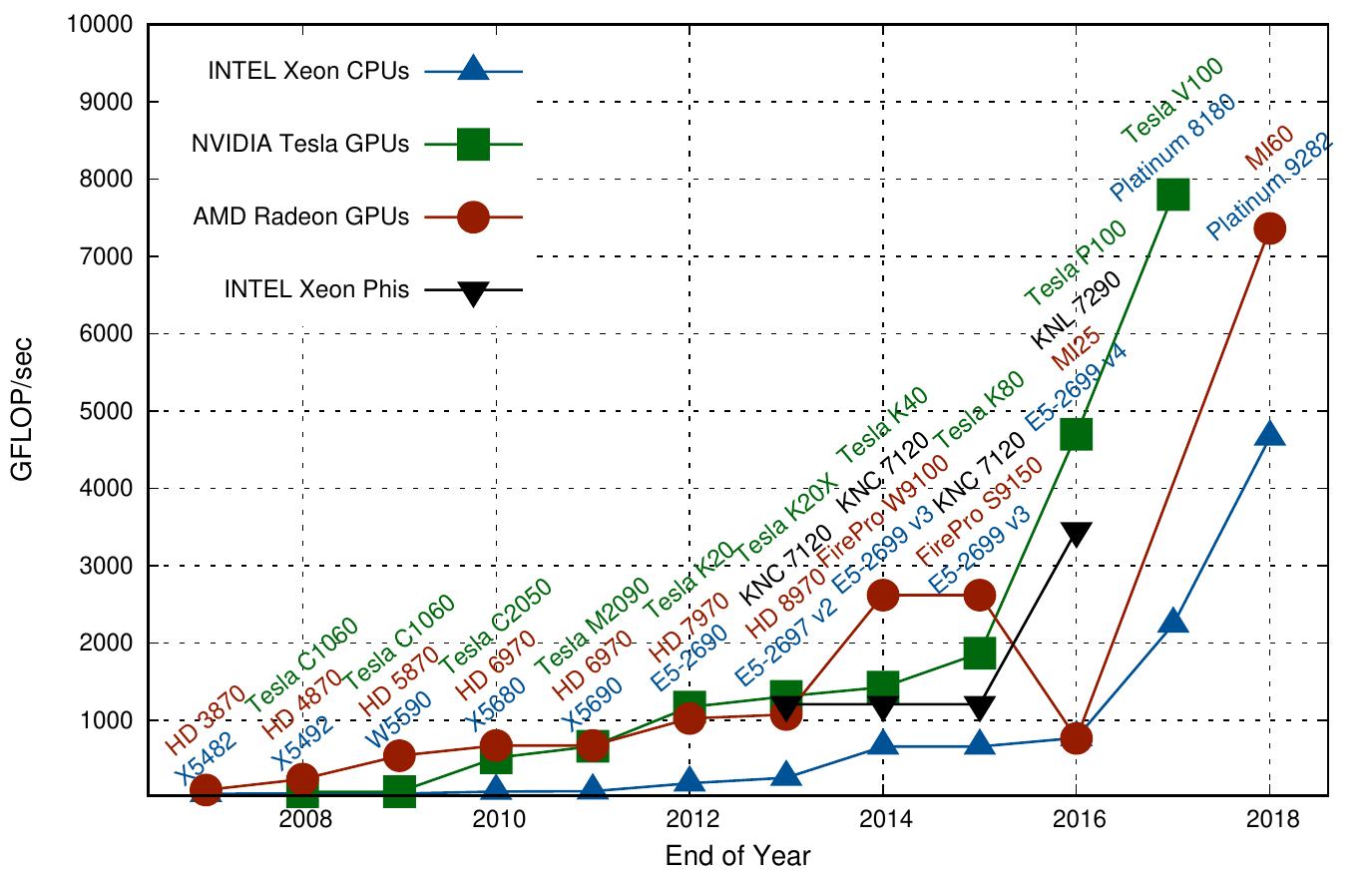}
\caption{Comparison of double-precision computing performance}
\label{fig:gpuvscpu}
\end{figure}

        The Graphics Processing Unit (GPU) is a specialized processor originally
designed specifically for accelerating video output with complex 3D scenes
in games, CADs and visual effects. In order to optimize the speed of
positioning and texturing of complex geometries (which involves
interpolation), the GPUs have become rich in floating-point compute
throughput. Furthermore, the GPU architecture has developed a massive
parallelism as a fastest approach to process millions of independent
triangles. In early 2000's, enthusiasts already turned their interest
towards general-purpose computations on GPUs (GPGPU) for science and
research. However, the GPU programming frameworks of that time (such as
GLSL shading language) implied mandatory video output, and did not offer
some important features, such as fast shared memory. The GPGPU market has
been revolutionized by NVIDIA Corporation in 2007 with the release of
GeForce 8 consumer graphic cards with both traditional video processing and
compute-only GPGPU pipelines. The GPGPU functionality has become widely
available to users through the CUDA v1.0 software development kit~\citep{nvidia2010programming}.
Also, NVIDIA released Tesla C870 - the first compute-dedicated graphics
card for data centers. Although early GPUs did not offer
double-precision computations support, their single-precision performance was a lot
higher than on CPUs of the same generaton. During over a
decade, NVIDIA GPU technology has seen multiple architectural updates
(Tesla, Fermi, Kepler, Maxwell, Volta). The latest NVIDIA Tesla V100 GPU
has double-precision peak compute throughput of 7.8 TFLOPS ~\citep{krupp} (see Fig.~\ref{fig:gpuvscpu}). The CUDA toolkit and
GPGPU ecosystem nowadays has grown into a large set of free de-facto
standard libraries and frameworks for different purposes: linear algebra
(CUBLAS, CUSPARSE), FFT (CUFFT), big data (Thrust), machine learning
(TensorFlow), etc.

In our phase tracking method, DE algorithm is employed for solving multidimensional optimization problem. The DE algorithm requires thousands of objective function evaluations, each integrating several dozen points of time series. Along with the basic double precision floating-point multiply-add, the computation of objective function invloves trigonometric operations. Given that the objective function integration is computationally expensive and parallelizable, it becomes an ideal candidate for GPU processing. We offload this part to the GPU by using the Thrust C++ framework~\citep{hoberock2010thrust}.

\begin{table}[!htbp]  \caption{CPU and GPU computing speed comparison}
\centering
{\scriptsize
\label{tab:gpuvscpu}
\begin{tabular}{rrrr}
\toprule
  data volume [Mbit]&  GTX580 (GPU) [ms]  & OpenMP (CPU) [ms]&VML (CPU) [ms]\\   
\midrule
    1  &   0.2  &    0.5  &  0.7 \\
    2  &   0.3  &    0.6  &  1.6 \\
    4  &   0.3  &    2.2  &  3.5 \\
    8  &   0.5  &    6.5  &  7.8 \\
   16  &   0.9  &   13.4  & 13.3 \\
   32  &   1.7  &   25.2  & 25.2 \\
   64  &   3.3  &   48.6  & 50.2 \\
  128  &   6.4  &   68.6  &100.7 \\
  256  &  12.6  &  107.3  &201.2 \\
\bottomrule
\end{tabular}
}
\end{table}
Table~\ref{tab:gpuvscpu} compares the time to solution on NVIDIA GTX 580 GPU and 2$\times$ Intel E5-2620v2 CPU (which have peak double precision performance of 196 and 60 GFLOPs, respectively). Two implementations have been evaluated on CPU: OpenMP and Intel VML (Vector Math library). The GPU is shown to give about 2$\times$ speedup against CPU for a small dataset (below 2M). The speedup improves significantly for larger datasets and reaches 9$\times$ for 256M. Furthermore, the OpenMP implementation is faster than VML. Table~\ref{tab:time} shows the GPU performance of a single objective function processing steps corresponding to Eq.~\ref{eq:2-2-2} for 1M dataset. 

\begin{table}[!htbp]  \caption{\centerline{ Sinlge objective function processing steps on NVIDIA GTX580 (1 Mbit dataset)}}
\centering
\label{tab:time}
\begin{tabular}{ll}
\toprule
manipulation &  time consumed \\
\midrule
Array initialization & 0.039ms \\
Elements square of array & 0.015ms \\
Array linear algebra operation & 0.017ms \\
Array trigonometric operation & 0.016ms \\
Array summation & 0.118ms \\
Total time$^a$  & 0.205ms \\
\bottomrule
\end{tabular}

\footnotesize{ $^a$ Without host$\leftrightarrow$GPU data transfer time}\\
\end{table}

In a real experiment, the data block size is about 400K (0.082ms on GTX580) and is evaluated by the DE solver $\sim$80,000 times to reach the convergence threshold (totals 6.5s on GTX580). Furthermore, by replacing GTX580 with 2$\times$ NVIDIA K80s (3.6 TFLOPs peak) we were able to acheive real-time processing.

\section{Receiver and signal model}
    \subsection{Receiver overview}
        The Planetary Radio Science Receiver (PSRS) has been developed under the contract between Shanghai Astronomical Observatory
and Southeast University of China and started operation in 2008. The design of PSRS is similar to the RSR receiver framework by
JPL~\citep{Sniffin2005dsms} with 70MHz IF signal input. PSRS has two output channels: the first channel is a narrow band signal 
recorded on a hard drive after two-stage down conversion from the IF signal; the second channel is a hardware Doppler output after one-stage down conversion. Table~\ref{tab:device} shows the available PRSR data recording modes.

\begin{table}[!htbp]  \caption{PRSR data recording modes}
\centering
\label{tab:device}
\resizebox{0.48\textwidth}{!}{  %调整字体
\begin{tabular}{|c|c|c|c|}
\hline
output signal & bandwidth/sampling &quantization & \tabincell{c}{data rate\\(bit/s)}\\
\hline
\multirow{8}*{narrow band} &\multirow{2}*{  16 KHz/32KHz } & 8 & 256,000\\
\cline{3-4} 
								&								 & 16 &512,000\\
\cline{2-4}
								&\multirow{2}*{  25 KHz/50KHz } & 8 & 400,000\\
\cline{3-4}
								&								 & 16 &800,000\\
\cline{2-4}
								&\multirow{2}*{  50 KHz/100KHz } & 8 & 800,000\\
\cline{3-4}
								&								 & 16 &1600,000\\
\cline{2-4}
								&\multirow{2}*{  100 KHz/200KHz } & 8 & 1600,000\\
\cline{3-4}
								&								 & 16 &3200,000\\
\hline
\multirow{4}*{wide band}   &  1 MHz/2MHz  & 1,2,4,8 & 16,000,000\\
\cline{2-4}
								&  2 MHz/4MHz  & 1,2,4,8 & 32,000,000\\
\cline{2-4}
								&  4 MHz/8MHz  & 1,2,4 & 32,000,000\\
\cline{2-4}
								&  8 MHz/16MHz  & 1,2 & 32,000,000\\
\hline
\end{tabular}
}
\end{table}

\textcolor{green}{
In addition to the PSRS receiver, we also use the CDAS(Chinese VLBI Data Acquisition System) receiving device developed by the Shanghai Observatory in the Change'E series missions. 
}

    \subsection{Signal model and objective function}
        The first PSRS channel recordered on hard disk has phase errors of various physical origins. The continuous data stream is processed in segments or \emph{blocks} of length $T$ (the value of T is related to the motion of the spacecraft, and is usually set to 2 seconds). If the relative timestamp in the center of the data block has the phase zero, then the whole block can be expanded by a finite Taylor polynomial within the interval [-T/2,T/2]:

\begin{equation}\label{eq:2-2-1}
\begin{split}
\phi(t)=\phi(0)+\sum^n_{i=1}\frac{\phi^i(0)}{i!}t^i
\end{split}
\end{equation}

Here $n$ is the order of Taylor polynomial which is also related to the motion of the spacecraft. In most of the cases $n=3$ meets the truncation error requirements when block length is set to $T=2$ seconds. Generally, the truncation error is set to the phase noise, which will be discussed later. With the Taylor polynomial phase expression, the radio signal in the data block can be expressed as:

\begin{equation}\label{eq:2-2-2}
\begin{split}
s(t)&=(c_4+c_5t)\cos(\phi(t))\\
    &=(c_4+c_5t)\cos(c_0+c_1t+c_2t^2+c_3t^3)
\end{split}
\end{equation}

Here, $c_{0\sim3}$ are the coefficients of Taylor polynomial:

\begin{equation}\label{eq:2-2-3}
\begin{split}
c_0&=\phi(0)\\
c_i&=\frac{\phi^i(0)}{i!}\big|_{i=1,2,3}
\end{split}
\end{equation}

The coefficients $c_4$ and $c_5$ are signal amplitude parameters. Moreover, $c_5$ is the slope of amplitude and it can be ignored when amplitude change is small with data length below 10 seconds.

Differential Evolution (DE) algorithm minimizes the objective function, which is generally non-negative. From Eq.~\ref{eq:2-2-2} we construct two types of objective functions. The first one is the classical $\chi^2$ function and the second one is based on the correlation function:

\begin{equation}\label{eq:2-2-4}
\begin{split}
F^{\chi^2}_{obj}(c_k)&=\sum^N_{i=1}\frac{[s_i-s(c_k,t_i)]^2}{\sigma^2_i}\Big|_{\sigma_i=1}\\
           &=\sum^N_{i=1}[s_i-s(c_k,t_i)]^2 
\end{split}
\end{equation}
\begin{equation}\label{eq:2-2-5}
\begin{split}
F^{cor}_{obj}(c_k)=\frac{1}{\arctan[{\rm cor}(s_i,s(c_k,t_i))]+\frac{\pi}{2} }
\end{split}
\end{equation}

In Eq.~\ref{eq:2-2-4}, $\frac{1}{\sigma^2_i}$ is the weighting factor that is relative to the standard deviation of
signal influenced by physical environment. Influence is equal to every data point. So weighting factor can be set to 1.
The optimal $c_k$ coefficients are to be determined. The $s(c_k,t_i)$ is the signal model constructed in
Eq.~\ref{eq:2-2-2}, and $s_i$ is the quantized signal value. The $F^{\chi^2}_{obj}(c_k)$ and $F^{cor}_{obj}(c_k)$ objective functions shall converge to the sum of noise power, and to $\frac{1}{\pi}\sim 0.318$, respectively. In Eq.~\ref{eq:2-2-5} the signal model is slightly changed, such that the amplitude of signal is constant; otherwise, the objective function becomes divergent:

\begin{equation}\label{eq:2-2-6}
\begin{split}
s(t) =A\cos(c_0+c_1t+c_2t^2+c_3t^3)
\end{split}
\end{equation}

\begin{figure}[!h]
\centering
\includegraphics[width=8cm,height=7cm]{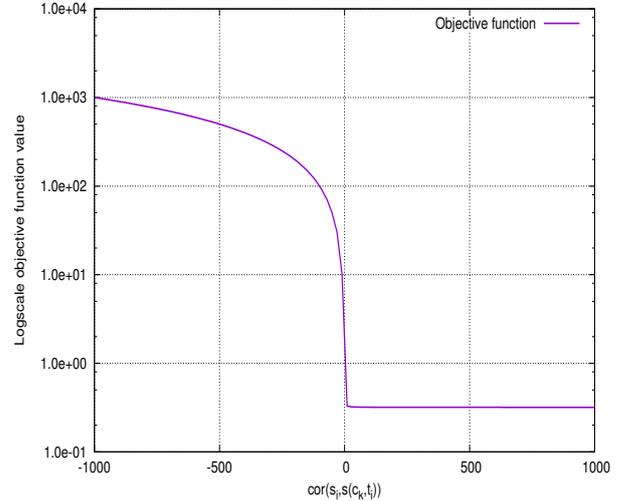}
\caption{Relation between the objective function and the correlation coefficient}
\label{fig:2-2-1}
\end{figure}

\begin{figure}[!h]
\centering
\includegraphics[width=9cm,height=6cm]{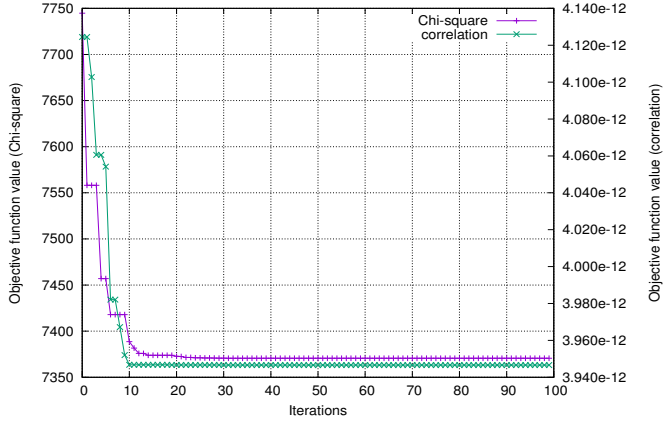}
\caption{Comparison of convergence rate of two objective functions types}
\label{fig:2-2-2}
\end{figure}

Here, $A$ is the average amplitude of signal. Fig.~\ref{fig:2-2-1} shows the relation between $F^{cor}_{obj}(c_k)$ and ${\rm cor}(s_i,s(c_k,t_i))$. The figure also shows that the convergence is very fast when the correlation coefficient is close to zero. Fig.~\ref{fig:2-2-2} compares the convergence rate of two objective functions types.

The data used for comparison is the real Mars Express (MEX) tracking data with the baseband sampling of 200KHz. \textcolor{red}{For the convenience of display, the correlation objective function value is subtracted a constant of 0.31830988618.} The figure shows the $F^{cor}_{obj}$ is faster to converge than $F^{\chi^2}_{obj}$, and takes only 10 times iterations. On the other hand, $F^{\chi^2}_{obj}$ requires 25 times iterations. Moreover, the computational difficulty of one $F^{cor}_{obj}$ evaluation is only 60\% of $F^{\chi^2}_{obj}$ function. In total, the $F^{cor}_{obj}$ takes 75\% less time to compute a data block, compared to $F^{\chi^2}_{obj}$. Although the $F^{cor}_{obj}$ is faster than $F^{\chi^2}_{obj}$, it cannot handle larger block sizes, which have significant signal amplitude change.

\section{Data processing and error analysis}
    \subsection{Data processing}
        This subsection will introduce the parameters setting for DE algorithm, phase continuity checking and process flow in real data processing.
\subsubsection{Determination of parameters range}
Determination range of parameters to be solved are required when calling DE algorithm which must cover the true value of parameters. 
Because DE algorithm is a kind of global optimisation, generally speaking range of parameters can be enlarged widely enough 
to cover the true value of parameters. But this strategy will cause huge amount of computation. So properly determine the initial value
and range of parameters is necessary for high efficient computation. The range Determination includes initial range setting
and running time range setting. Following content of this subsection will discuss these two kind of range settings based on
$\chi^2$ function model mentioned in Eq.~\ref{eq:2-2-4}.

Determination of proper initial range of parameters needs estimation of parameters. From Eq.~\ref{eq:2-2-3} we can see
that the polynomial coefficients are determined by derivatives of phase at block center. So if we can estimate the 
derivatives of phase at block center then polynomial coefficients could be properly determined. The detailed estimation
process is as follows:
\begin{enumerate}[noitemsep,topsep=0pt]
\it
\small
\item $c_0$ is the instantaneous phase at block center, range can be set as $[0,2\pi]$.
\item $c_1$ is the instantaneous frequency at block center (\SI{}{\radian\per\second}). It can be determined through FFT computation
of a small piece of data extracted from block center. Range of $c_1$ can be setted as $\pm$10\% of $c_1$.
\item $c_2$ is equal to half of second derivative of phase at block center. So the value of $c_2$ can be estimated by
central difference method through three equally spaced frequency values near block center. The frequency value ls calculated as step 2 
\item $c_3$ is the 1/6 of third derivative of phase at block center. This term is very small and will not be greater
than 10 in all situations. In real data processing initial range of $c_3$ is set to $[-50,50]$ for the sake of insurance.
\item $c_4$ is the amplitude of signal. It can be determined by FFT algorithm done at step 2. Range of $c_4$ can be
set as $\pm$10\% of $c_4$.
\item $c_5$ is the slope of amplitude which relative uncertainty is big when block length is not longer enough. For the sake of
insurance it's range is set as $[-0.2c_4\frac{1}{T},0.2c_4\frac{1}{T}]$.
\end{enumerate}
If there is a forecast orbit of spacecraft derivatives of phase at block center also can be properly estimated from orbit using
Eq.~\ref{eq:3-2-3}. 

Determination of range of parameters at running time is exactly automatic tracking of the signal. Extrapolation is used
in coefficients of Taylor polynomial estimation from $n$th block to $n+1$th block. The specific running time range estimation of  is as follows:
\begin{enumerate}[noitemsep,topsep=0pt]
\it
\small
\item Range of $c_0^{n+1}$ is the same as initial range setting $[0,2\pi]$
\item Value of $c_1^{n+1}$ can be extrapolated from the first-order phase derivative of block $n$ (Eq.~\ref{eq:2-2-1}): 
    $\dot{\phi}^n(t)=c_1^{n}+2c_2^{n}t+3c_3^{n}t^2$. So the value is $c_1^{n+1}=\dot{\phi}^n(T)$, and range could be set
        as $[\dot{\phi}^n(T/2),\dot{\phi}^n(3T/2)]$\footnote{The sign will be decided by software automatically.}
    \item Value of $c_2^{n+1}$ can be extrapolated as $c_1^{n+1}$, that is  $c_2^{n+1}=0.5\ddot{\phi}^n(T)$. Range could be set 
    as $[0.5\ddot{\phi}^n(T/2),0.5\ddot{\phi}^n(3T/2)]$
\item Value of $c_3$ is small and the range can be further narrowed from the result of first time of DE calling.
\item Range of $c_4^{n+1}$ is $[c_4^{n}-0.1c_4^{n},c_4^{n}+0.1c_4^{n}]$
\item Range of $c_5$ can be fixed as $c_3$
\end{enumerate}
Range setings by means of extrapolation of $c_1^{n+1}$ and $c_2^{n+1}$ are effective in most situations that is when derivatives of phase is
monotonic functions. But when derivatives of phase are not monotonic functions within two blocks the range of parameters
will not cover the true value and parameters searching will failed. This generally happens while the spacecraft crossing the orbital 
apogee or perigee. The software will automatically restart searching after properly enlarge the range of parameters.
Meanwhile the values of $c_3$ and $c_5$ are random distribution when data length is not long enough(that is the signal to noise ratio
is not big enough)(~\ref{sec:app1})

\subsubsection{The DE algorithm control parameters setting}
The DE algorithm runs under a set of control parameters. Table~\ref{tab:3-1-1} gives the explanation of these control
parameters and values set for running.

\begin{table}[!htbp]  \caption{DE algorithm control parameters}
    \centering
    \label{tab:3-1-1}
    \scriptsize
    \begin{tabular}{lll}
        \toprule
        item  &   Meaning  &  value \\
        \midrule
        $\rm Dim\_XC$ &  dimension of problem  & 6\\
        XCmin$^a$ & The lower bound of parameters & *\\
        XCman$^a$ & The upper bound of parameters & *\\
        VTR   & The expected fitness value of objective function & 0 \\
        itermax & The maximum number of iterations &  200 \\
        $\rm F\_XC$ &Mutation scaling factor & 0.5 \\
        $\rm CR\_XC$ & Crossover factor  & 0.85\\
        strategy & The strategy of the mutation operations& 3\big|6 \\
        \bottomrule
    \end{tabular}

\footnotesize{ $^a$ Items of parameters range }\\
\end{table}
There are six kind of mutation operations strategies in the DE algorithm which affect the solution robustness and 
computation speed\citep{price2006differential}. 
Among these strategies the third is the best choice for objective 
function Eq.~\ref{eq:2-2-4} and the sixth for Eq.~\ref{eq:2-2-5}. Two type of stop condition are generally used for parameters searching.
The first is fixing the iteration number when proper parameters range can be determined at running time. The second is
setting the threshold of change of objective function or parameters. In practical applications parameters and objective
function sometimes remain the same in several iterations because of the characteristics of genetic algorithms. So 
the first type of stop condition is used in real data processing.   

\subsubsection{Phase continuity checking }
A natural method to judge weather solution is perfect is to check the continuity of instantaneous phase and frequency at
the bound of two adjacent data block. Using Eq.~\ref{eq:2-2-1} we can get the expression of phase and frequency:
\begin{equation}\label{eq:3-1-1}
\begin{split}
\phi(t)=&c_0+c_1t+c_2t^2+c_3t^3\\
{\rm F}(t)=&\frac{d\phi(t)}{dt}\\
        =&c_1+2c_2t+3c_3t^2
\end{split}
\end{equation}
Phase and frequency at the bound of two adjacent data block (n and n+1) are:
\begin{equation}\label{eq:3-1-2}
\small
\begin{split}
\phi^{n}|_{right}=&c^n_0+c^n_1t+c^n_2t^2+c^n_3t^3|_{t=T/2} \pmod {2\pi}\\
\phi^{n+1}|_{left}  =&c^{n+1}_0+c^{n+1}_1t+c^{n+1}_2t^2+\\
             &c^{n+1}_3t^3|_{t=-T/2} \pmod {2\pi}\\
{\rm F}^n|_{right}=&c^n_1+2c^n_2t+3c^n_3t^2|_{t=T/2}\\
{\rm F}^{n+1}|_{left}  =&c^{n+1}_1+2c^{n+1}_2t+3c^{n+1}_3t^2|_{t=-T/2}\\
\end{split}
\end{equation}
Checking weather the values of $\phi^{n}|_{right}$ and ${\rm F}^n|_{right}$ are equal to $\phi^{n+1}|_{left}$ and ${\rm
F}^{n+1}|_{left}$ within the threshold of noise we can confirm the correctness of parameters result from DE algorithm. 
A real data processing example of MEX shows continuity of phase and frequency in ~\ref{sec:app1}. The noise of phase 
is about \SI{50} {\milli\radian} (1 $\sigma$, 1 second integration) in MEX tracking. So the continuity of phase can be confirmed.

\subsubsection{Data process flow }
\begin{figure}[!h]
\centering
\includegraphics[width=8.4cm,height=10cm]{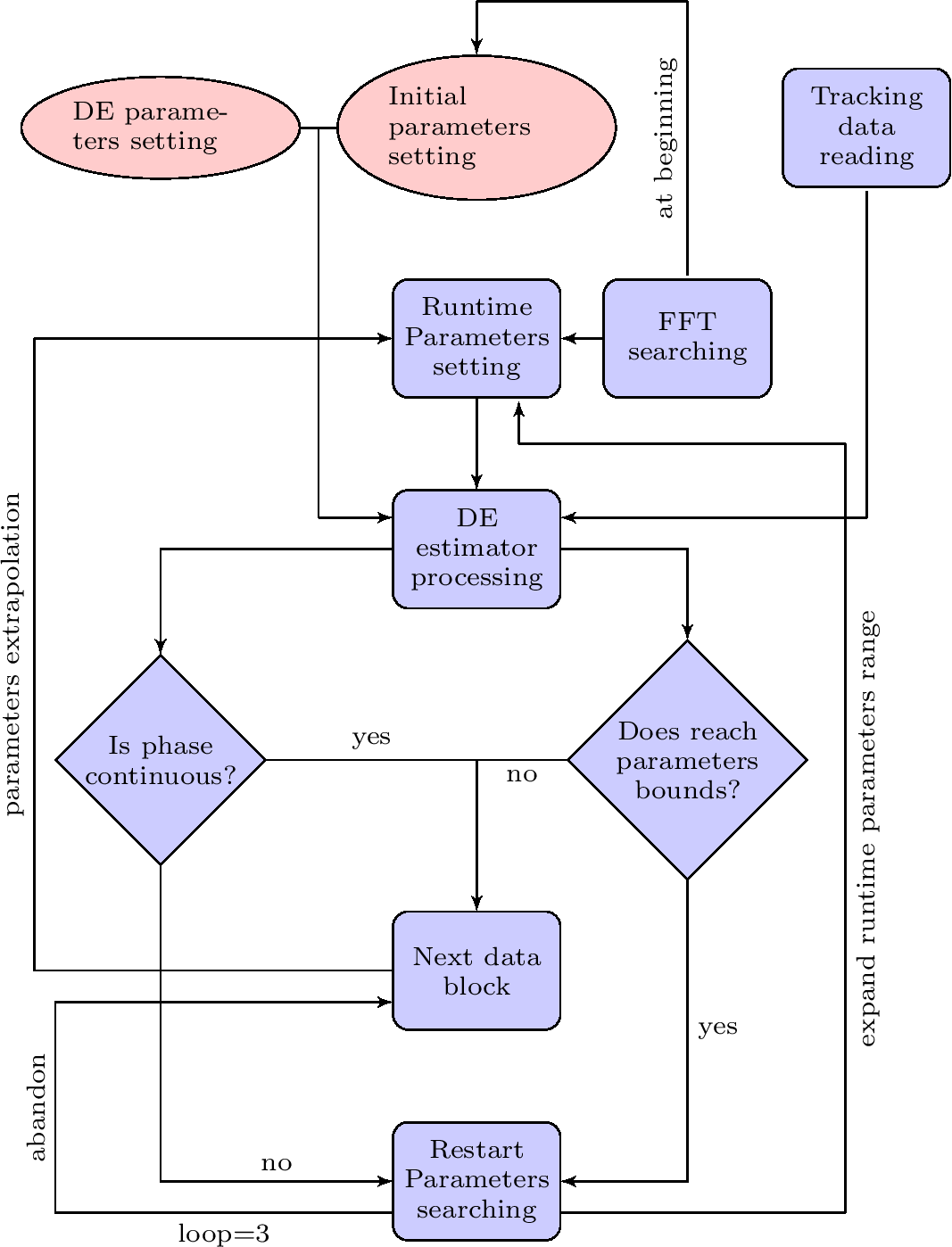}
\caption{Data process flow}
\label{fig:3-1-1}
\end{figure}

The software reads the initial parameters range and the DE algorithm controls parameters from control file and runs 
automatically without manual intervention. Fig.~\ref{fig:3-1-1} gives the process flow.
In radio measurement there exist some accidental factors that cause phase distortion and iterations divergence. In this situation the
software will automatically adjust parameters range and try to search again. The data block will be abandoned after three 
attempts.

    \subsection{Error analysis}
        There are three kind of errors in data process. The first error is the truncation error in phase calculation which is
also called remainder of Taylor polynomial. The second error is random errors in tracking data. The third is system error
caused by equipments and medias along the signal path. Remainder of Taylor polynomial is relative to the 
block length and polynomial order. Random errors come from thermal noise of signal beacon, receiver and medias along
signal path. System error can be eliminated by construction of error models in high level data process which is no in
the scope of this paper and will not be discussed. 
\subsubsection{Remainder of Taylor polynomial}
Remainder of Taylor polynomial has several expression forms. Lagrange error bound is usually used for error analysis.
Lagrange error bound of Eq.~\ref{eq:2-2-1} can be written as:
\begin{equation}\label{eq:3-2-1}
\begin{split}
R_n(t)=\frac{\phi^{n+1}(\xi)}{(n+1)!}t^{n+1}\Big|_{\xi\in[0,t]}
\end{split}
\end{equation}
$\xi$ is a point within [0,t]. The max error within data block is generally at the block border. Remainder at 
block border is:
\begin{equation}\label{eq:3-2-2}
\begin{split}
R_n(\frac{T}{2})&=\frac{1}{2^{n+1}(n+1)!}\phi^{(n+1)}(\xi)T^{n+1}\Big|_{\xi\in[0,\frac{T}{2}]}\\
                &< \frac{1}{2^{n+1}(n+1)!}\big|\phi^{(n+1)}_{max}\big|T^{n+1}
\end{split}
\end{equation}
Here $\big|\phi^{(n+1)}_{max}\big|$ is maximum  absolute value of $n+1$th derivative of phase within data block.
From Eq.~\ref{eq:3-2-2} we can see the upper bound of remainder is relative to the order of Taylor polynomial $n$, length of data block $T$
and the absolute value of maximum $n+1$th derivative of phase. Generally speaking 
\footnote{In the following contents of this paper, if not specified, the default value of $n=3$.} $n=3$ can ensure the remainder 
small enough with which upper bound of $\big|\phi^{(n+1)}_{max}\big|$ will
be \SI[parse-numbers = false, number-math-rm = \ensuremath]{3\times 10^{-7}}{\radian\per\s\tothe{4}} (See ~\ref{sec:app2}). When the block length is not so long(less than 10 seconds) in this situation the upper bound 
of truncation error of Taylor polynomial is \SI[parse-numbers = false, number-math-rm = \ensuremath]{10^{-5}}{\radian}. But in some special cases such as aircraft passing across
perigee or flying-by a giant planet amplitude of $\big|\phi^{(n+1)}_{max}\big||_{n=3}$ can not be ignored (this
term is about \SI[parse-numbers = false, number-math-rm = \ensuremath]{2\times 10^{-3}}{\radian\per\s\tothe{4}} while MEX passing across perigee) and  $n=3$ is not enough for high precision data 
processing. In this situation $n=4$ will be work and the upper bound of truncation error could be controlled within
\SI[parse-numbers = false, number-math-rm = \ensuremath]{10^{-4}}{\radian}.
Derivatives of phase can be estimated from the derivatives of line-of-sight velocity using forecast orbit:
\begin{equation}\label{eq:3-2-3}
\begin{aligned}
\big|\phi^{(n+1)}_{max}\big|\simeq& \frac{2\pi  f_0}{c}\big|v^{(n)}_{max}\big|  & \qquad (1-way)\\
\big|\phi^{(n+1)}_{max}\big|\simeq& \frac{4\pi M_2f_0}{c}\big|v^{(n)}_{max}\big|  & \qquad (3-way)
\end{aligned}
\end{equation}

In theory, if there is no effect of thermal noise the truncation error can be controlled very small by adjusting the
parameters $n$ and $T$. But thermal noise has traits that cannot be eliminated. Excessively reducing the truncation
error is futile. So the threshold of truncation error can be set to the thermal noise of phase $\sigma_{noise}(T)$. Here 
$T$ is integration time for measuring phase noise which is exactly the block length. P{\"a}tzold et al. have given a relative 
expression of $\sigma_{noise}(T)$
\begin{equation}\label{eq:3-2-4}
\begin{split}
\sigma_{noise}(T)=\sigma_{noise}(1)\sqrt{\frac{1}{T}}
\end{split}
\end{equation}
$\sigma_{noise}(1)$ is standard thermal noise when integration time is equal to 1 second. From the Eq.~\ref{eq:3-2-4} we can infer 
that thermal noise can be effectively suppressed by increasing the integration time.

\begin{figure}[tbp]
\centering
\includegraphics[width=8.4cm,height=6cm]{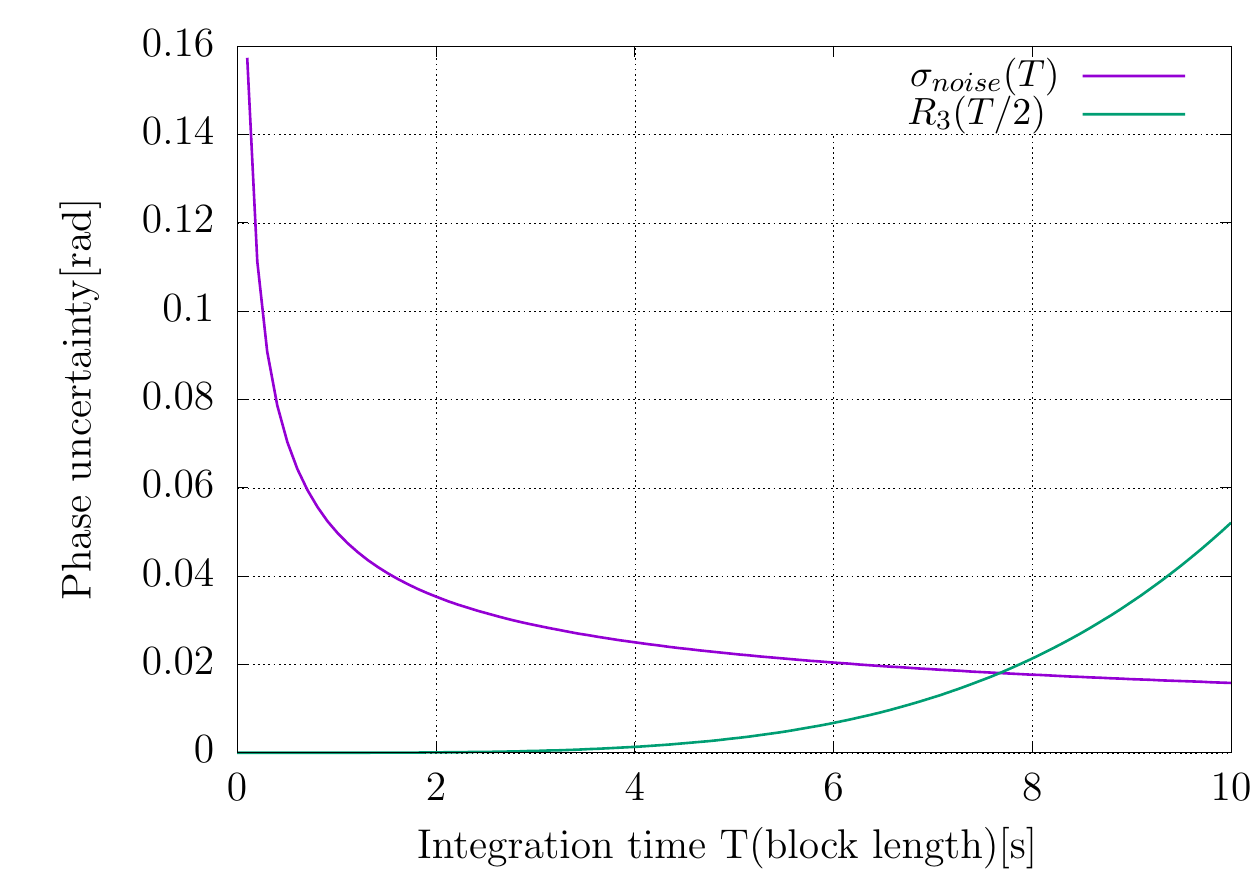}
\caption{Variety of Phase noise $\sigma_{noise}(T)$ and remainder $R_{3}(\frac{T}{2})$}
\label{fig:3-2-1}
\end{figure}

There exists a choice of $T$ to balance the truncation error and thermal noise from Eq.~\ref{eq:3-2-2} and
Eq.~\ref{eq:3-2-4}. On the one hand, the remainder and $T$ are positively correlated. On the other hand, the thermal noise
and $T$ are negatively correlated. The length of $T$ is worth considering to balance the computation amount and the data 
process precision. Fig.~\ref{fig:3-2-1} gives the relationship of truncation error and phase thermal noise
relative to block length in MEX tracking case (\SI[parse-numbers = false, number-math-rm = \ensuremath]{\sigma_{noise}(1) = 50}{\milli\radian}).
From Fig.~\ref{fig:3-2-1} we can see that truncation error will be less than phase thermal noise when block length is shorter than 7
seconds. In the actual data processing, to ensure that the truncation error is less than the phase noise and improve the 
data processing efficiency, the data block length is generally set to 2 seconds.

\subsubsection{Random phase noise}
Random noise in data will affect the estimation of phase and it can be used as criteria for truncation error as
described above. Random noise comes from two sources. The first is the thermal noise of signal source, transponder on board and receiver at
station. The second is random interference of solar phase scintillations along signal path. Carrier phase error variance 
can be expressed as~\citep{Sniffin2005dsms}:   
\begin{equation}\label{eq:3-2-5}
\begin{split}
\sigma^2_{noise}|_{1-way}&=\sigma^2_{S}+\frac{1}{\rho_{L}}\\
\sigma^2_{noise}|_{3-way}&=\frac{M^2(B_{TR}-B_L)}{P_C/N_0|_{U/L}}+\sigma^2_{S}+\frac{1}{\rho_{L}}
\end{split}
\end{equation}
Here,
\begin{itemize}[noitemsep,topsep=0pt]
\small
\it
\item $\sigma^2_{S}$ is contribution to carrier loop phase error variance due to solar phase scintillations
\item $\rho_{L}$ is downlink carrier loop signal-to-noise ratio
\item $M$ is transponder ratio
\item $B_{TR}$ is one-sided, noise-equivalent, transponder carrier loop bandwidth
\item $B_L$ is one-sided, noise-equivalent, loop bandwidth of downlink carrier loop
\item $P_C/N_0|_{U/L}$ is uplink carrier power to noise spectral density ratio
\end{itemize}

Phase error variance contributed by solar phase scintillations is related to the angle SEP (Sun Earth Probe).
This effect is greater when the sun is near the direction of observation. Eq.~\ref{eq:3-2-6} gives the relation between
$\sigma^2_{S}$ and SEP~\citep{Sniffin2005dsms}:
\begin{equation}\label{eq:3-2-6}
\begin{split}
\sigma^2_{S}=\frac{C_{band}C_{loop}}{({\rm sin}\theta_{SEP})^{2.45}B_L^{1.65}}\big|_{\theta_{SEP}\in[\SI{5}{\degree},\SI{27}{\degree}]}
\end{split}
\end{equation}
In Eq.~\ref{eq:3-2-6} $C_{band},C_{loop}$ are constants which vary with frequency band. For X band of up and down link
$C_{band}=1.9\times 10^{-6},C_{loop}=5.9$. 

The typical magnitudes of phase noise for 2/3 way tracking at X band
are~\citep{yuen2013deep}:
\begin{equation}\label{eq:noiseP}
\sigma^2_{noise}\le \begin{cases}
\SI{0.1} {\radian\square} &  {\rm residual\, carrier}\\
\SI{0.02} {\radian\square} & {\rm suppressed\, carrier\, BPSK}\\
\SI{0.005} {\radian\square} & {\rm QPSK}
\end{cases}
\end{equation}
Eq.~\ref{eq:noiseP} gives the upper bound of phase noise variance for three types of carrier modulation.

\section{Applications}
        Using the Taylor polynomial coefficients estimated by the DE algorithm we can form several kinds of observables including integration Doppler, instantaneous
Doppler, total count phase and line-of-sight acceleration.

    \subsection{Doppler observables}
        \subsubsection{Integration Doppler}
Integration Doppler is an important observation type in planetary gravity research. The original 
definition of integration Doppler is the phase change during an fixed count interval~\citep{moyer2005formulation}:
\begin{equation}\label{eq:4-1-1}
\begin{split}
F_{int\_dop}=\frac{\phi_{t_e}-\phi_{t_s}}{T_c}
\end{split}
\end{equation}
Here we ignore the bias for the convenience of discussion. $t_e$ is the station time at the end of count interval and
$t_s$ is the station time at the beginning of count interval. $T_c$ is the count interval length which is different for different 
 research object. Typical count times have durations of tens of seconds in planetary gravity research while spacecraft
 orbiting planet. In testing theory of general relativity and gravitational wave
 searching~\citep{asmar1993deep,asmar2009planetary} interval of counting will be few thousand seconds for extremely high
 Doppler measurement precision during interplanetary cruise. 
 
The typical data block length is 2 seconds and the limitation of data block length is 20
seconds limited by the processing ability of the GPU. We can use phase connection to form the longer integration Doppler
observables:
\begin{equation}\label{eq:4-1-2}
\begin{split}
\Phi_{total}&=\sum^N_{i=1}\big[\phi_{i}(\frac{T}{2})-\phi_{i}(-\frac{T}{2})\big]\Big|_{N=\frac{T_{c}}{T}}\\
F_{int\_dop}&=\frac{\Phi_{total}}{T_{c}}
\end{split}
\end{equation}
In Eq.~\ref{eq:4-1-2} $i$ is serial number of block, $\phi_{i}(\frac{T}{2})$ can be evaluated by phase expression
(Eq.~\ref{eq:3-1-1}), $\Phi_{total}$ is the phase change within count interval $T_c$ which is also a new kind of observation type 
in planetary science research~\citep{moyer2005formulation}. The column 8 of ~\ref{sec:app1} gives the example of
$[\phi_{i}(\frac{T}{2})-\phi_{i}(-\frac{T}{2})]$. Uncertainty of Integration Doppler can be expressed as:
\begin{equation}\label{eq:4-1-3}
\begin{split}
\sigma^2_{int\_dop}=\frac{\sigma^2(T)}{NT}
\end{split}
\end{equation}
Meaning of the Eq.~\ref{eq:4-1-3} is the same as Eq.~\ref{eq:3-2-4}. $\sigma^2(T)$ is phase variance of block or the phase discontinuous
variance at block border. 

\begin{figure}[tbp]
\centering
\includegraphics[width=8.4cm,height=6cm]{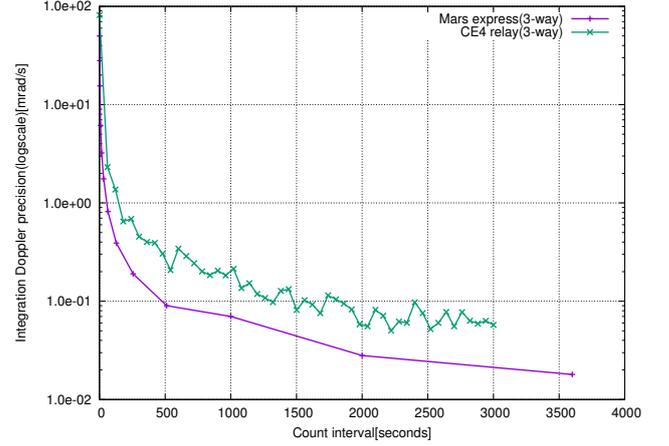}
\caption{Precision of integration Doppler changes with count interval}
\label{fig:4-1-1}
\end{figure}

Figure~\ref{fig:4-1-1} shows the precision of integration Doppler with different interval scales in MEX and CE4 data process.
The Doppler precision is about \SI{50}{\milli\radian\per\second}: for MEX with 1 second Integration schale and \SI{80}{\milli\radian\per\second}: for 
CE4 relay satellite.

\subsubsection{Instantaneous Doppler}
Instantaneous Doppler is used in planetary occultation research in which high Doppler resolution is required at ingress
or egress of radio signal into or out of planetary atmosphere. There are two kinds of approximate approaches to compute
instantaneous Doppler. The first is to decrease the count interval~\citep{moyer2005formulation} and use integration Doppler 
to approximate instantaneous Doppler. The second is using short time Fast Fourier Transform (FFT) to compute an average frequency of a short
data block~\citep{patzold2000mars}. The Doppler precision of these two approaches will obey Eq.~\ref{eq:3-2-4} that is
$3.16$ times of $\sigma_{noise}(1)$ when Doppler sampling rate is 10 Hz. 

Using Eq.~\ref{eq:3-1-1} we can conveniently calculate instantaneous Doppler at any time tag within data block and 
the precision is guaranteed by the truncated error $\sigma_{noise}(2)$ which is 0.7 times of $\sigma_{noise}(1)$. 
So instantaneous Doppler calculated from Taylor polynomial has higher precision than method of short count interval or
method of short time FFT.

    \subsection{Line-of-sight acceleration observables}
        From Eq.~\ref{eq:3-2-3} we can see that derivatives of phase is related with the derivatives of line-sight-of velocity
and second derivative of phase can be approximated as function of line-sight-of acceleration:
\begin{equation}\label{eq:4-2-1}
\begin{aligned}
\phi^{(2)}\simeq& \frac{2\pi  f_0}{c}a_{los}  & \qquad (1-way)\\
\phi^{(2)}\simeq& \frac{4\pi M_2f_0}{c}a_{los}  & \qquad (3-way)
\end{aligned}
\end{equation}
A strict and full expansion of Eq.~\ref{eq:4-2-1} can be found in section ~\ref{sec:app3} which gives connection
between observable of $\phi^{(2)}$ and dynamic state of spacecraft and stations. Combined with a tracking case of MEX
magnitude analysis of line-of-sight acceleration is given in which magnitude of line-of-sight acceleration in Newtonian 
frame is about 1 $m/s^2$, cross velocity terms $\delta^i_{ij}$ have magnitude of $10^{-3}m/s^2$ in first order of
Eq.~\ref{eq:app3-9}, and main term of second order of Eq.~\ref{eq:app3-9} is about $10^{-4}m/s^2$. 

Base on the concept of ~\ref{sec:app3} we give an estimation of the mass and two order of gravity of
Phobos~\citep{jian2019new}. The line-of-sight acceleration observable ($\phi^{(2)}$) directly reflects the kinematic state of the
spacecraft ($\mathbf{r}_2,\dot{\mathbf{r}}_{2},\ddot{\mathbf{r}}_{2}$) and 
can be used to estimate the dynamic parameters relative to spacecraft($\ddot{\mathbf{r}}_{2}$) corporation with a reference orbit.
It also can be used in process of least square regress combined with traditional Doppler observables to solve
parameters.

\section{Conclusion}
        This paper introduces a new phase tracking method for planetary radio science research based on GPGPU (General
Purpose Computing on GPU) technology. In addition to Doppler observables used in planetary radio science research
the new method also gives the line-of-sight acceleration observable which directly reflects the dynamic state of the spacecraft.
Using line-of-sight acceleration we can directly solve the dynamic parameters interested with a reference orbit.

%\section{Conclusion}
%        \input{files/file6}
\section{Acknowledgments}
        This work has been supported by the National Natural Science Foundation of China, the Astronomical Joint Program (U1531104). We thank Mr. Gou Wei from Sheshan 25-meter Observatory for assistance in MEX tracking, Prof. Meng Qiao, Chen Congyan and Lu Rongjun of Southeast University of China for advice on Doppler data processing, and Prof. Ping Jinsong of National Observatory of China for helpful conversations.

%Bibliography
%\bibliographystyle{elsart-harv} % (For numbered Elsevier citations)
\bibliography{files/publications}

\onecolumn
\begin{appendix}
    \section{1}\label{sec:app1} 
    {\bfseries Phase tracking output example for MEX tracking }
\begin{scriptsize}
\begin{verbatim}
               UTC time             c_0(rad) c_1(rad/s)    c_2(rad/s^2)    c_3      c_4          c_5     total_phase   RSS       data quality index 
      (border) 2013-12-28T19:24:38   6.102   146782.529
block1(center) 2013-12-28T19:24:39   5.670   146566.029    -108.151807     0.066    4728.3       6.7     293132.189    3424636.172    1
      (border) 2013-12-28T19:24:40   2.563   146349.922
      (border) 2013-12-28T19:24:40   2.585   146349.721
block2(center) 2013-12-28T19:24:41   3.319   146133.984    -107.928933    -0.040    4713.0     -18.4     292267.887    3417344.713    1
      (border) 2013-12-28T19:24:42   1.824   145918.005
      (border) 2013-12-28T19:24:42   1.831   145918.155
block3(center) 2013-12-28T19:24:43   4.412   145702.273    -107.838436     0.068    4721.2     -10.5     291404.683    3425031.927    1
      (border) 2013-12-28T19:24:44   4.946   145486.801
               2013-12-28T19:24:44   4.933   145486.642
...            2013-12-28T19:24:45   3.474   145271.151    -107.703009     0.028    4749.4      -2.9     290542.359    3427429.823    1
               2013-12-28T19:24:46   0.237   145055.830
               2013-12-28T19:24:46   0.238   145055.809
...            2013-12-28T19:24:47   1.568   144840.493    -107.600678     0.038    4710.2       2.9     289681.063    3423038.571    1
               2013-12-28T19:24:48   1.325   144625.406
               2013-12-28T19:24:48   1.316   144625.163
               2013-12-28T19:24:49   5.865   144410.414    -107.439969    -0.044    4725.1     -35.5     288820.741    3419870.188    1
               2013-12-28T19:24:50   2.878   144195.403
               2013-12-28T19:24:50   2.893   144195.292
               2013-12-28T19:24:51   4.936   143980.762    -107.331095    -0.044    4726.0      -0.9     287961.436    3424449.594    1
               2013-12-28T19:24:52   5.946   143765.968
               2013-12-28T19:24:52   5.962   143766.015
               2013-12-28T19:24:53   5.951   143551.478    -107.244130     0.016    4711.1      -4.7     287102.989    3424489.034    1
               2013-12-28T19:24:54   5.081   143337.038
               2013-12-28T19:24:54   5.106   143337.164
               2013-12-28T19:24:55   3.608   143122.853    -107.113158     0.028    4710.5      -5.7     286245.762    3419445.809    1
               2013-12-28T19:24:56   1.512   142908.711
               2013-12-28T19:24:56   1.497   142908.715
               2013-12-28T19:24:57   5.237   142694.705    -106.952688     0.035    4709.9       2.9     285389.480    3418848.526    1
               2013-12-28T19:24:58   2.417   142480.904
               2013-12-28T19:24:58   2.437   142480.624
               2013-12-28T19:24:59   5.570   142267.018    -106.825240    -0.015    4682.0       6.7     284534.007    3433902.863    1
               2013-12-28T19:25:00   2.398   142053.323
...
...


Here, integration length of total_phase (colum 8 above) is the block length 
\end{verbatim}
\end{scriptsize}

    \section{2}\label{sec:app2} 
    \begin{table}[!htbp]  \caption{Related parameters of the error remainder estimation of MEX and Cassini}
\centering
\label{tab:erf1}
{\footnotesize
\begin{threeparttable}[t]
\begin{tabular}{p{2.5cm}cccllll}
\specialrule{0em}{2pt}{2pt}
\toprule
\multirow{2}*{Arcs/band} &\multirow{2}*{$\sigma_{noise}$[\si{\milli\radian}]\tnote{$\gamma$} }&\multirow{2}*{\tabincell{c}{ Polynomia\\ order}} &\multirow{2}*{ Block length[${\rm s}$]}&
\big|$\phi^{(2)}_{max}$\big|[\si{\radian\per\square\second}]& \big|$\phi^{(3)}_{max}$\big|[\si{\radian\per\s\tothe{3}}] &
\big|$\phi^{(4)}_{max}$\big|[\si{\radian\per\s\tothe{4}}]      & \big|$\phi^{(5)}_{max}$\big|[\si{\radian\per\s\tothe{5}}]  \\
 \cline{5-8} & & & &
\big|$a_{max}$\big|[\si{\meter\per\square\second}] &\big|$a^{(1)}_{max}$\big|[\si{\meter\per\second\tothe{3}}]&\big|$a^{(2)}_{max}$\big|[\si{\meter\per\second\tothe{4}}]&\big|$a^{(3)}_{max}$\big|[\si{\meter\per\second\tothe{5}}]\\
\midrule

\multirow{2}*{ \tnote{$\alpha$}\quad Mex(cruise)/X(2AU) }  &\multirow{2}*{ 50 }   &\multirow{2}*{ 2}  &\multirow{2}*{ 10} &  
 $7\times 10^{0}$    &$5\times 10^{-4}$ & $3\times 10^{-7}$ &   \\
\cline{5-8} & & & &
\tnote{$\epsilon$}\, $3\times 10^{-2}$ &$1\times 10^{-6}$ &$1\times 10^{-10}$ & \\
\hline
\multirow{2}*{ \tnote{$\alpha$}\quad Mex(orbiting)/X} &\multirow{2}*{ 50} &\multirow{2}*{4\big|3\tnote{$\delta$}} &\multirow{2}*{2}   &
$8\times 10^{2}$    &$6\times 10^{-1}$ & $2\times 10^{-3}$ & $3\times 10^{-6}$   \\
\cline{5-8} & & & &
 $2\times 10^{0}$ &$2\times 10^{-3}$ &$6\times 10^{-6}$ &$8\times 10^{-9}$  \\
\hline
\multirow{2}*{ \tnote{$\beta$}\quad Cas(cruise)/X(10AU) }& \multirow{2}*{ 20}    & \multirow{2}*{ 2}  & \multirow{2}*{ 10}   & 
$7\times 10^{0}$    &$5\times 10^{-4}$ & $3\times 10^{-7}$ &     \\
\cline{5-8} & & & &
\tnote{$\epsilon$}\, $3\times 10^{-2}$ &$1\times 10^{-6}$ &$1\times 10^{-10}$ & \\
\hline
\multirow{2}*{ \tnote{$\beta$}\quad Cas(orbiting)/X} &\multirow{2}*{ 20 } &\multirow{2}*{ 4\big|3\tnote{$\delta$}}  &\multirow{2}*{ 2}   & 
$3\times 10^{3}$    &$2\times 10^{0}$  & $3\times 10^{-3}$ & $6\times 10^{-6}$  \\
\cline{5-8} & & & &
 $8\times 10^{0}$ &$6\times 10^{-3}$ &$8\times 10^{-6}$ & $2\times 10^{-8}$\\
\bottomrule
\end{tabular}
     \begin{tablenotes}
     \item[$\alpha$] 3-way tracking.
     \item[$\beta$]  3-way tracking.
     \item[$\gamma$] One second integration time.
     \item[$\delta$] 4 order when near perigee, 3 order with normal arcs
	 \item[$\epsilon$] The main contribution is the rotation of the earth
\end{tablenotes}
\end{threeparttable}
}
\end{table}

    \section{3}\label{sec:app3} 
    {\bfseries Observation function of second derivative of phase}\\

Acronyms:
\begin{itemize}[noitemsep,topsep=0pt] %[topsep=0pt,itemsep=0pt,parsep=0pt,before=\vspace{1cm},after=\vspace{1mm}]
\small
\item 1 \ \ foot mark means uplink station
\item 2 \ \ foot mark means satellite 
\item 3 \ \ foot mark means downlink station
\item c \ \ speed of light in vacuum
\item $t_{i}$ \ \ coordinate time at position i
\item $\mathbf{r}_{i}$ \ \ position vector of object i in inertia frame
\item $\mathbf{r}_{ij}$ \ \ position vector of object j relative to i
\item $\dot{\mathbf{r}}_{i}$ \ \ velocity vector of object i in inertia frame
\item $\dot{\mathbf{r}}_{ij}$ \ \ velocity vector of object j to i
\item $\ddot{\mathbf{r}}_{i}$ \ \ acceleration vector of object i in inertia frame
\item $F_T$ \ \ frequency transmitted from uplink station 
\item $M_2$ \ \ transponder turnaround ratio onboard
\item $F_R$ \ \ frequency received by downlink station
\item $\tau_3$ \ \ atomic clock of downlink station
\item $r_{ij}$ \ \ distance from position i to j
\item $\dot{r}_{ij}$ \ \ deviation of distance from position i to j
\item $\Phi_{i}$ \ \ Newtown gravitation at position i  
\item $\dot{s}_{i}$ \ \ velocity magnitude of object i relative to Sun   
\item SSB \ \ Solar System Barycenter
\item $L_B=1.550520^{-8}$ \ \ Lagrangian constant of astronomy 
\end{itemize}

\subsubsection{2/3 way tracking model}
Theodore D. Moyer gives a precise formula accurate to $1/c^2$ describing the sky frequency received by downlink station in 2/3-way tracking 
models~\citep{moyer1971mathematical}.
\begin{equation}\label{eq:app3-1}
\begin{split}
(1-\frac{F_R}{M_2F_T})=\frac{1}{c}(\dot{r}_{12}+\dot{r}_{23})+\frac{1}{c^2}\left[\dot{r}_{12}\dot{p}_{12}+\dot{r}_{23}\dot{p}_{23}-\dot{r}_{12}\dot{r}_{23}+(\phi_1-\phi_3)+ \frac{1}{2}(\dot{s}^2_1-\dot{s}^2_3)\right]
\end{split}
\end{equation}
Using above equation the derivative of $F_R$ to $\tau_3$ can be written as($1/c$ terms):
\begin{equation}\label{eq:app3-2}
\begin{split}
    \diff1{F_R}{\tau_3}\Big|_{o(1/c)}=-\frac{M_2F_T}{c}(\frac{d\dot{r}_{12}}{d\tau_3}+\frac{d\dot{r}_{23}}{d\tau_3})
\end{split}
\end{equation}
Expansion of right side of above equation:
\begin{equation}\label{eq:app3-3}
\begin{split}
\frac{d\dot{r}_{12}}{d\tau_3}&=\opartial{\dot{r}_{12}}{t_1}\diff1{t_1}{t_2}\diff1{t_2}{t_3}\diff1{t_3}{\tau_3} + \opartial{\dot{r}_{12}}{t_2}\diff1{t_2}{t_3}\diff1{t_3}{\tau_3}\\
\frac{d\dot{r}_{23}}{d\tau_3}&=\opartial{\dot{r}_{23}}{t_2}\diff1{t_2}{t_3}\diff1{t_3}{\tau_3} + \opartial{\dot{r}_{23}}{t_3}\diff1{t_3}{\tau_3}\\
\end{split}
\end{equation}
Expansion of partial derivatives of velocity to coordinate time:
\begin{equation}\label{eq:app3-4}
\begin{split}
\opartial{\dot{r}_{12}}{t_1}&=+\frac{1}{r^3_{12}}(\mathbf{r}_{12}\cdot\dot{\mathbf{r}}_{1})(\mathbf{r}_{12}\cdot\dot{\mathbf{r}}_{12})
                              -\frac{1}{r_{12}}(\dot{\mathbf{r}}_{1}\cdot\dot{\mathbf{r}}_{12}+\mathbf{r}_{12}\cdot\ddot{\mathbf{r}}_{1})\\
\opartial{\dot{r}_{12}}{t_2}&=-\frac{1}{r^3_{12}}(\mathbf{r}_{12}\cdot\dot{\mathbf{r}}_{2})(\mathbf{r}_{12}\cdot\dot{\mathbf{r}}_{12})
                              +\frac{1}{r_{12}}(\dot{\mathbf{r}}_{2}\cdot\dot{\mathbf{r}}_{12}+\mathbf{r}_{12}\cdot\ddot{\mathbf{r}}_{2})\\
\opartial{\dot{r}_{23}}{t_2}&=+\frac{1}{r^3_{23}}(\mathbf{r}_{23}\cdot\dot{\mathbf{r}}_{2})(\mathbf{r}_{23}\cdot\dot{\mathbf{r}}_{23})
                              -\frac{1}{r_{23}}(\dot{\mathbf{r}}_{2}\cdot\dot{\mathbf{r}}_{23}+\mathbf{r}_{23}\cdot\ddot{\mathbf{r}}_{2})\\
\opartial{\dot{r}_{23}}{t_3}&=-\frac{1}{r^3_{23}}(\mathbf{r}_{23}\cdot\dot{\mathbf{r}}_{3})(\mathbf{r}_{23}\cdot\dot{\mathbf{r}}_{23})
                              +\frac{1}{r_{23}}(\dot{\mathbf{r}}_{3}\cdot\dot{\mathbf{r}}_{23}+\mathbf{r}_{23}\cdot\ddot{\mathbf{r}}_{3})
\end{split}
\end{equation}
Relation between coordinate time and station time~\citep{moyer1971mathematical}:
\begin{equation}\label{eq:app3-5}
\begin{split}
\frac{dt_3}{d\tau_3}=&(1-L_B)(1-\frac{2\Phi_3}{c^2}-\frac{\dot{s_3}^2}{c^2})^{-\frac{1}{2}}\approx 1+\frac{\Phi_3}{c^2}+\frac{\dot{s_3}^2}{2c^2}-L_B\\
\end{split}
\end{equation}
Relation of coordinate time of different objects and definitions of relative velocity:
\begin{equation}\label{eq:app3-6}
\begin{split}
\frac{dt_1}{dt_2}&=1-\frac{\dot{r}_{12}}{c}-\frac{\dot{r}_{12}\dot{p}_{12}}{c^2}\\
\frac{dt_2}{dt_3}&=1-\frac{\dot{r}_{23}}{c}-\frac{\dot{r}_{23}\dot{p}_{23}}{c^2}\\
\frac{dt_1}{dt_2}\frac{dt_2}{dt_3}&=1-\frac{\dot{r}_{12}+\dot{r}_{23}}{c}+\frac{\dot{r}_{12}\dot{r}_{23}-\dot{r}_{12}\dot{p}_{12}-\dot{r}_{23}\dot{p}_{23}}{c^2}\\
            \dot{r}_{12}&=\frac{\mathbf{r}_{12}}{r_{12}}\cdot\dot{\mathbf{r}}_{12}\\
            \dot{r}_{23}&=\frac{\mathbf{r}_{23}}{r_{23}}\cdot\dot{\mathbf{r}}_{23}\\
            \dot{p}_{12}&=\frac{\mathbf{r}_{12}}{r_{12}}\cdot\dot{\mathbf{r}}_1\\
            \dot{p}_{23}&=\frac{\mathbf{r}_{23}}{r_{23}}\cdot\dot{\mathbf{r}}_2\\
			\dot{\mathbf{r}}_{12}&=\dot{\mathbf{r}}_{2}-\dot{\mathbf{r}}_{1}\\
			\dot{\mathbf{r}}_{23}&=\dot{\mathbf{r}}_{3}-\dot{\mathbf{r}}_{2}\\
\end{split}
\end{equation}
By substituting Eq.~\ref{eq:app3-2},~\ref{eq:app3-3},~\ref{eq:app3-4},~\ref{eq:app3-5},~\ref{eq:app3-6} into
Eq.~\ref{eq:app3-1} we can get the first order of $1/c$ expansion for the derivative of $F_R$ to $\tau_3$.

Expansion of $1/c^2$ terms of Eq.~\ref{eq:app3-1}\footnote{The last two terms are very small($\simeq 10^{-10}$) and will be
ignored.}:
\begin{equation}\label{eq:app3-7}
\begin{split}
\diff1{F_R}{\tau_3}\Big|_{o(1/c^2)}&=-\frac{M_2F_T}{c^2}\left[\diff1{(\dot{r}_{12}\dot{p}_{12})}{\tau_3}+\diff1{(\dot{r}_{23}\dot{p}_{23})}{\tau_3}-\diff1{(\dot{r}_{12}\dot{r}_{23})}{\tau_3} \right]\\
       &=-\frac{M_2F_T}{c^2}\left[\opartial{\dot{r}_{12}}{\tau_3}\dot{p}_{12}+ \opartial{\dot{p}_{12}}{\tau_3}\dot{r}_{12}+ 
                                   \opartial{\dot{r}_{23}}{\tau_3}\dot{p}_{23}+ \opartial{\dot{p}_{23}}{\tau_3}\dot{r}_{23}-
                                   \opartial{\dot{r}_{12}}{\tau_3}\dot{r}_{23}- \opartial{\dot{r}_{23}}{\tau_3}\dot{r}_{12}\right]
\end{split}
\end{equation}
Some terms have been expanded in Eq.~\ref{eq:app3-3} and~\ref{eq:app3-6} and remains can be expanded as:
\begin{equation}\label{eq:app3-8}
\begin{split}
\opartial{\dot{p}_{12}}{\tau_3}&=\opartial{\dot{p}_{12}}{t_1}\diff1{t_1}{t_2}\diff1{t_2}{t_3}\diff1{t_3}{\tau_3} + \opartial{\dot{p}_{12}}{t_2}\diff1{t_2}{t_3}\diff1{t_3}{\tau_3}\\
\opartial{\dot{p}_{23}}{\tau_3}&=\opartial{\dot{p}_{23}}{t_2}\diff1{t_2}{t_3}\diff1{t_3}{\tau_3} + \opartial{\dot{p}_{23}}{t_3}\diff1{t_3}{\tau_3}\\
\end{split}
\end{equation}
Derivatives of different time scales can be found in Eq.~\ref{eq:app3-5} and~\ref{eq:app3-6}. Remain terms of
Eq.~\ref{eq:app3-8} can be expanded as:
\begin{equation}\label{eq:app3-9}
\begin{split}
\opartial{\dot{p}_{12}}{t_1}&=+\frac{1}{r^3_{12}}(\mathbf{r}_{12}\cdot\dot{\mathbf{r}}_{1})(\mathbf{r}_{12}\cdot\dot{\mathbf{r}}_{1})
                              -\frac{1}{r_{12}}(\dot{\mathbf{r}}_{1}\cdot\dot{\mathbf{r}}_{1}-\mathbf{r}_{12}\cdot\ddot{\mathbf{r}}_{1})\\
\opartial{\dot{p}_{12}}{t_2}&=-\frac{1}{r^3_{12}}(\mathbf{r}_{12}\cdot\dot{\mathbf{r}}_{2})(\mathbf{r}_{12}\cdot\dot{\mathbf{r}}_{1})
                              +\frac{1}{r_{12}}(\dot{\mathbf{r}}_{2}\cdot\dot{\mathbf{r}}_{1})\\
\opartial{\dot{p}_{23}}{t_2}&=+\frac{1}{r^3_{23}}(\mathbf{r}_{23}\cdot\dot{\mathbf{r}}_{2})(\mathbf{r}_{23}\cdot\dot{\mathbf{r}}_{2})
                              -\frac{1}{r_{23}}(\dot{\mathbf{r}}_{2}\cdot\dot{\mathbf{r}}_{2}-\mathbf{r}_{23}\cdot\ddot{\mathbf{r}}_{2})\\
\opartial{\dot{p}_{23}}{t_3}&=-\frac{1}{r^3_{23}}(\mathbf{r}_{23}\cdot\dot{\mathbf{r}}_{3})(\mathbf{r}_{23}\cdot\dot{\mathbf{r}}_{2})
                              +\frac{1}{r_{23}}(\dot{\mathbf{r}}_{3}\cdot\dot{\mathbf{r}}_{2})\\
\end{split}
\end{equation}
By substituting Eq.~\ref{eq:app3-5},~\ref{eq:app3-6},~\ref{eq:app3-8},~\ref{eq:app3-9} into Eq.~\ref{eq:app3-7} we can get
the expansion of $1/c^2$ terms of Eq.~\ref{eq:app3-1}

Formal the second derivative of phase can be written as:
\begin{equation}\label{eq:app3-9}
\begin{split}
\diffi{\phi}{\tau_3}{2}=\diff1{F_R}{\tau_3}(\mathbf{r}_i,\dot{\mathbf{r}}_{i},\ddot{\mathbf{r}}_{i})\Big|_{o(1/c)}+\diff1{F_R}{\tau_3}(\mathbf{r}_i,\dot{\mathbf{r}}_{i},\ddot{\mathbf{r}}_{i})\Big|_{o(1/c^2)}
\end{split}
\end{equation}
Here scale of station time $\tau_3$ is the same as time tag of data block. So the phase derivatives are equal to
each other:
\begin{equation}\label{eq:app3-10}
\begin{split}
\diffi{\phi}{\tau_3}{2}=\diffi{\phi}{t}{2}
\end{split}
\end{equation}

\subsubsection{Magnitude estimation of $\diffi{\phi}{\tau_3}{2}$ with case of MEX mission}
Considering a common 3-way tracking of MEX with uplink station of New Norcia located at Australia and downlink station of
Sheshan located at China:
\begin{itemize}[noitemsep,topsep=0pt] 
\small 
\item Distance from station to MEX: \ \ $r_{12}\simeq r_{23}=2\times 10^8$ Km
\item Distance between stations:\ \ $r_{13}\simeq3000$ Km
\item Velocity of station relative to SSB: \ \ $\dot{r}_{1}\simeq \dot{r}_3=30 $ Km/s
\item Velocity of MEX relative to SSB \ \ $\dot{r}_2\simeq 30$Km/s
\end{itemize}
Considering that the distance between station and spacecraft is much larger than the distance between stations, the following
relationship is defined and established.
\begin{equation}\label{eq:app3-11}
\begin{split}
\hat{\mathbf{e}}&=\frac{\mathbf{r}_{23}}{r_{23}}\\
                &\simeq -\frac{\mathbf{r}_{12}}{r_{12}}
\end{split}
\end{equation}
Here, $\hat{\mathbf{e}}$ is unit vector of opposite direction of line-of-sight. 

\vskip 0.8cm
{\small {\bfseries First order of Eq.~\ref{eq:app3-9}}}\\
By ignoring the influence of gravitation
on time scale derivative(Eq.~\ref{eq:app3-5}) first order of Eq.~\ref{eq:app3-9} can be simplified as:
\begin{equation}\label{eq:app3-12}
\begin{split}
 \diff1{F_R}{\tau_3}(\mathbf{r}_i,\dot{\mathbf{r}}_{i},\ddot{\mathbf{r}}_{i})\Big|_{o(1/c)} \simeq-\frac{M_2F_T}{c}\Big[&(\delta^1_{12}+\hat{\mathbf{e}}\cdot\ddot{\mathbf{r}}_{1})(1-\beta_v)+\\
                                                            &(\delta^2_{12}+\delta^2_{23}-2\hat{\mathbf{e}}\cdot\ddot{\mathbf{r}}_{2})(1-\beta_u)+\\
                                                            &(\delta^3_{23}+\hat{\mathbf{e}}\cdot\ddot{\mathbf{r}}_{3}) \Big]\\
\end{split}
\end{equation}
In the equation Amplitude of $\ddot{\mathbf{r}}_{1}$ and $\ddot{\mathbf{r}}_{3}$ in the frame of SSB is
about $10^{-2}m/s^2$ due to Earth rotation and amplitude of $\ddot{\mathbf{r}}_{2}$(MEX)
is about $1m/s^2$ due to orbit motion. $\delta^i_{ij}$ are cross velocity terms with dimension of acceleration and 
with amplitude of $10^{-3}m/s^2$:
\begin{equation}\label{eq:app3-13}
\begin{split}
\delta^1_{12}&=+\frac{1}{r^3_{12}}(\mathbf{r}_{12}\cdot\dot{\mathbf{r}}_{1})(\mathbf{r}_{12}\cdot\dot{\mathbf{r}}_{12})
                              -\frac{1}{r_{12}}(\dot{\mathbf{r}}_{1}\cdot\dot{\mathbf{r}}_{12})\\
\delta^2_{12}&=-\frac{1}{r^3_{12}}(\mathbf{r}_{12}\cdot\dot{\mathbf{r}}_{2})(\mathbf{r}_{12}\cdot\dot{\mathbf{r}}_{12})
                              +\frac{1}{r_{12}}(\dot{\mathbf{r}}_{2}\cdot\dot{\mathbf{r}}_{12})\\
\delta^2_{23}&=+\frac{1}{r^3_{23}}(\mathbf{r}_{23}\cdot\dot{\mathbf{r}}_{2})(\mathbf{r}_{23}\cdot\dot{\mathbf{r}}_{23})
                              -\frac{1}{r_{23}}(\dot{\mathbf{r}}_{2}\cdot\dot{\mathbf{r}}_{23})\\
\delta^3_{23}&=-\frac{1}{r^3_{23}}(\mathbf{r}_{23}\cdot\dot{\mathbf{r}}_{3})(\mathbf{r}_{23}\cdot\dot{\mathbf{r}}_{23})
                              +\frac{1}{r_{23}}(\dot{\mathbf{r}}_{3}\cdot\dot{\mathbf{r}}_{23})
\end{split}
\end{equation}
Definition of $\beta$ factor(amplitude is about $10^{-4}$):
\begin{equation}\label{eq:app3-14}
\begin{split}
\beta_{u}&=\frac{\dot{r}_{23}}{c}\\
\beta_{v}&=\frac{\dot{r}_{12}+\dot{r}_{23}}{c}\\
\end{split}
\end{equation}
If we ignore the contributions of $\delta^i_{ij}$ and $\beta$, Eq.~\ref{eq:app3-12} can be further simplified 
as(in Newtonian frame):
\begin{equation}\label{eq:app3-15}
\begin{split}
 \diff1{F_R}{\tau_3}(\mathbf{r}_i,\dot{\mathbf{r}}_{i},\ddot{\mathbf{r}}_{i})\Big|_{o(1/c)} \simeq-\frac{M_2F_T}{c}[\hat{\mathbf{e}}\cdot\ddot{\mathbf{r}}_{1}-
                                                            2\hat{\mathbf{e}}\cdot\ddot{\mathbf{r}}_{2}+
                                                            \hat{\mathbf{e}}\cdot\ddot{\mathbf{r}}_{3} ]
\end{split}
\end{equation}
Here we can see that in Newton frame the second derivative of phase could be approximated as function of line-of-sight acceleration 
of $\ddot{\mathbf{r}}_{1},\ddot{\mathbf{r}}_{2}$ and $\ddot{\mathbf{r}}_{3}$ with error of $10^{-3}m/s^2$.     

\vskip 0.8cm
{\small {\bfseries Second order of Eq.~\ref{eq:app3-9}}}\\
The second order\footnote{Here $\frac{M_2F_T}{c}$ is treated as constant.} of Eq.~\ref{eq:app3-9} can be simplified to order of $\beta^2$:
\begin{equation}\label{eq:app3-16}
\begin{split}
    \diff1{F_R}{\tau_3}(\mathbf{r}_i,\dot{\mathbf{r}}_{i},\ddot{\mathbf{r}}_{i})\Big|_{o(1/c^2)}
\simeq-\frac{M_2F_T}{c}\{ 
             (1-\beta_v)&[(\beta_m+\beta_w)\hat{\mathbf{e}}\cdot\ddot{\mathbf{r}}_{1}+\beta_m\delta^1_{12}+\beta_w\delta^{12}_1 ] +\\
             (1-\beta_u)&[-(\beta_m+\beta_n+\beta_u)\hat{\mathbf{e}}\cdot\ddot{\mathbf{r}}_{2} +\beta_m\delta^2_{12}+\beta_n\delta^{2}_{23}+\beta_u\delta^{23}_2]+\\
             &[\beta_n\hat{\mathbf{e}}\cdot\ddot{\mathbf{r}}_{3}+\beta_n\delta^3_{23}] +
             \beta_w(1-\beta_u)\delta^{12}_2+\beta_u\delta^{23}_3  \}
\end{split}
\end{equation}
Here, $\delta^i_{ij}$ terms are defined in Eq.~\ref{eq:app3-13} and new definitions of $\beta$ are(with amplitude of $10^{-4}$):
\begin{equation}\label{eq:app3-17}
\begin{split}
\beta_{w}&=\frac{\dot{r}_{12}}{c}\\
\beta_{m}&=\frac{\dot{p}_{12}-\dot{r}_{23}}{c}\\
\beta_{n}&=\frac{\dot{p}_{23}-\dot{r}_{12}}{c}\\
\beta_{u,v,w,m,n} &\simeq 10^{-4}
\end{split}
\end{equation}
The main term of Eq.~\ref{eq:app3-16} approximates to $\beta\simeq10^{-4}m/s^2$.

\end{appendix}

\end{document}